# *Title:* *Particle–Hole Creation in Condensed Matter: A Conceptual Framework for Modeling Money–Debt Dynamics in Economics*

**Bumned Soodchomshom**

Department of physics, Faculty of science, Kasetsart University, Bangkok 10900, Thailand

E-mail: fscibns@ku.ac.th; bumned@hotmail.com

**Abstract**

We propose a field-theoretic framework that models money–debt dynamics in economic systems through a direct analogy to particle–hole creation in condensed matter physics. In this formulation, issuing credit generates a symmetric pair—money as a particle-like excitation and debt as its hole-like counterpart—embedded within a monetary vacuum field. The model is formalized via a second-quantized Hamiltonian that incorporates time-dependent perturbations to represent real-world effects such as interest and profit, which drive asymmetry and systemic imbalance. This framework successfully captures both macroeconomic phenomena, including quantitative easing (QE) and gold-backed monetary regimes, and microeconomic credit creation, under a unified quantum-like formalism. In particular, QE is interpreted as generating entangled-like pairs of currency and bonds, exhibiting systemic correlations akin to nonlocal quantum interactions. Asset-backed systems, on the other hand, are modeled as coherent superpositions that collapse upon use. This approach provides physicists with a rigorous and intuitive toolset to analyze economic behavior using many-body theory, laying the groundwork for a new class of models in econophysics and interdisciplinary field analysis.

**Keywords:** Econophysics; Money–Debt Dynamics; Particle–Hole Analogy; Second Quantization; Quantitative Easing (QE);

**Author Note:**

This work is intended as a conceptual and formal framework to bridge physical theories and monetary economics. Rather than providing empirical or numerical validation, our aim is to establish a rigorous analogy that enables future development of simulation models, data-driven formulations, or quantitative testing within econophysics. The model is not meant to replace classical economic theories but to complement them by offering physicists a language and structure to describe financial dynamics. We welcome interdisciplinary dialogue and extensions in either direction.

## 1. Introduction

In modern economic systems, the creation and circulation of money and debt lie at the core of financial dynamics. Yet, existing theories often address quantitative easing (QE), gold-backed currencies, and individual borrowing as discrete mechanisms. QE allows central banks to create new money by purchasing government bonds or other securities [1,2], while gold-standard regimes historically relied on physical assets to back currency issuance [3]. At the microeconomic level, personal credit is issued through banking systems based on collateral or future income expectations [4,5]. Although classical and modern frameworks—such as monetarism [6], Keynesianism [7], and endogenous money theory [2,5,8]—have sought to explain how money enters circulation and influences inflation, interest rates, and growth, they typically lack a unifying physical analogy to describe the symmetry, imbalance, and dynamic evolution inherent in monetary creation.

This paper introduces such a physical analogy by drawing on principles from quantum field theory and condensed matter physics. In particular, second quantization and many-body theory describe how particle–hole pairs emerge from excitations in a filled Fermi sea, where symmetry is preserved at equilibrium but may be broken under the influence of external perturbations [9–12]. We extend this analogy to economics: the issuance of credit is modeled as the creation of a symmetric money–debt pair—akin to particle–hole excitations in a monetary vacuum field. Within this framework, money plays the role of a mobile quasiparticle, while debt behaves as its complementary quasihole, confined within institutional or contractual structures. This approach allows for the formulation of a second-quantized Hamiltonian that captures both symmetry-preserving and symmetry-breaking behaviors. Time-dependent perturbations—such as interest rates and profits—shift the energy landscape, resulting in dynamic economic asymmetries. The model accommodates macroeconomic phenomena, including QE-induced entangled money–bond pairs and gold-backed monetary superpositions, as well as localized microeconomic credit events.

Ultimately, this framework seeks to unify the treatment of monetary operations across scales and systems. It provides a novel lens for physicists to understand economic behavior using field-theoretic methods, and for economists to explore monetary structures through the language of quantum systems and symmetry dynamics.

## 2. Theoretical Framework: Money/Debt as Particle/Hole Analogy

### 2.1 Field Analogy Between Physical and Economic Systems

To establish a foundational analogy, we treat the economic environment as a kind of "economic vacuum field," analogous to the physical vacuum in condensed matter systems. In physics, particle–hole pairs are created when an electron is excited from the valence band to the conduction band, leaving behind a hole. These are symmetric, energy-conserving events that occur in a vacuum-like medium and are fundamental to charge transport phenomena [9,10]. The process is depicted in Figure 1a:

- On the left, **electron–hole creation** illustrates an electron moving up to the conduction level, leaving a hole below.

- On the right, **recombination** shows the electron falling back and annihilating with the hole, releasing energy across the band gap.

We draw a direct economic analogy: when credit is issued, **money** is injected into the economy and begins circulating (analogous to the electron), while an equivalent **debt** is recorded in the credit system (analogous to the hole). This forms a symmetric money–debt pair at the moment of creation. The debt remains stationary in accounting systems, just as the hole remains localized in the valence band [2,5]. This concept is illustrated in **Figure 1b**:

- The **money–debt pair creation** mimics excitation, with money entering the liquidity level and debt anchored at the obligation level.

- The **recombination** shows repayment: money returns to the institution, canceling the debt. This collapse resembles annihilation of the exciton [9,11].

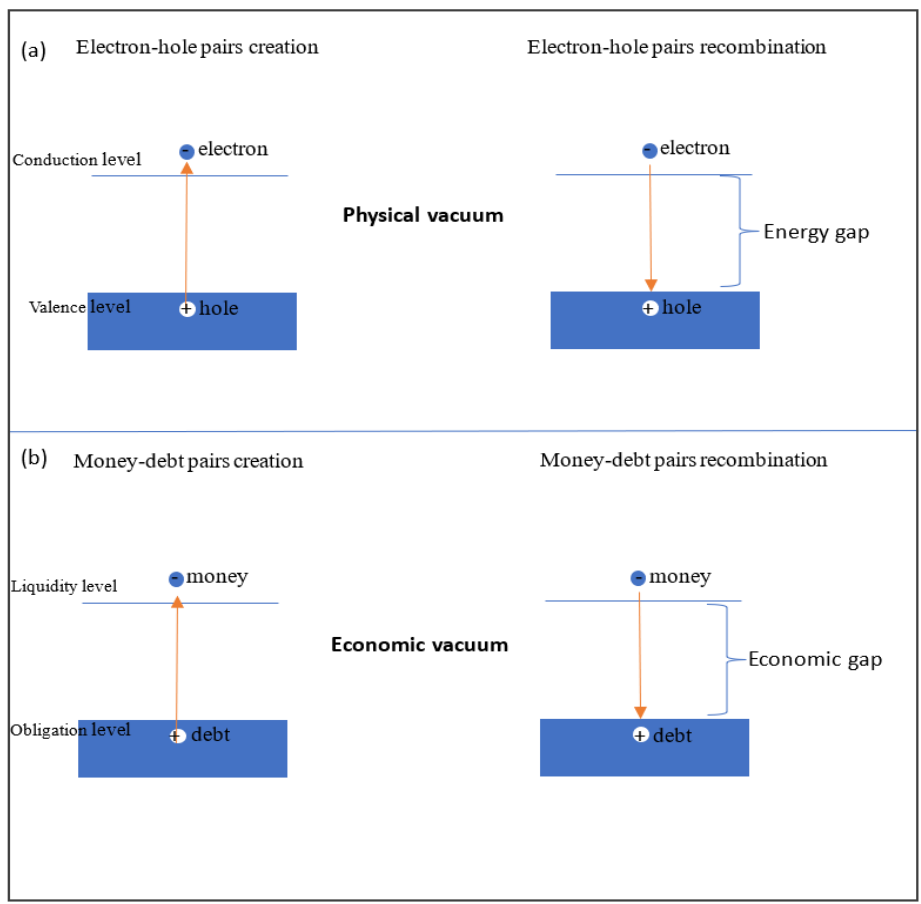

**Figure 1: Analogy between electron–hole pair creation in a physical vacuum and money–debt pair creation in an economic vacuum.** (a) In condensed matter systems, an electron is excited from the valence band to the conduction band, leaving behind a hole. This electron–hole pair exists across an energy gap and may recombine, releasing energy. (b) In economic systems, the issuance of credit generates a money–debt pair: money enters circulation (liquidity level), while debt remains on institutional balance sheets (obligation level). This pair exists across an economic gap and may recombine upon repayment. The diagram highlights the structural symmetry and energy-like interpretation shared between physical and financial pair dynamics.

The vertical separation in both diagrams represents the **gap**—in physics, the **energy gap**, and in economics, the **economic gap**. This economic gap captures the tension between liquidity (access to usable money) and obligation (the liability to repay), which introduces systemic energy in the form of financial risk and time-evolving interest or profit [1,2,5]. By visualizing these levels as fields and the particles as excitations, we provide a basis to construct Hamiltonians and dynamic models for economic transactions, capturing not only creation but also recombination and transformation [10,11].

**This analogy can be made more explicit: the process of electron–hole pair creation is conceptually equivalent to taking out a loan.** In both cases, a symmetric pair is generated—an electron and a hole in physics, and money and debt in economics. Just as the electron moves freely in the conduction band while the hole remains as a record of absence in the valence band, money flows into the economy while the debt remains registered on institutional balance sheets [4,5]. **Likewise, recombination of the electron–hole pair—where the electron falls back and annihilates with the hole—is directly analogous to the repayment of debt.** When repayment occurs, money is removed from circulation and the corresponding debt is extinguished, restoring the system to a neutral state [2]. This highlights the structural symmetry of the model: creation corresponds to credit issuance, and recombination corresponds to debt settlement. By emphasizing this pair structure and its temporal evolution, we gain a dynamic picture of credit systems as excitonic processes within a structured monetary field.

## 2.2 Mathematical Representation of Credit-Based Pair Creation

Just as excitations in a Fermi Sea generate symmetric particle–hole pairs [9,11], economic systems generate money–debt pairs through actions like lending and credit issuance. In condensed matter physics, these excitations are treated using second quantization [9-11], where a particle is created in the conduction band and a corresponding hole is left in the valence band. Similarly, when a loan is issued in the economic system, new money is introduced (analogous to the particle), while a corresponding debt (analogous to the hole) is formed in the credit ledger.

In economic terms, the particle–hole pair represents a credit event: money is disbursed into the market and begins circulating, while the debt is retained as a stationary liability recorded by the lending institution. The system is neutral at the moment of pair creation. In the non-interacting case, the Hamiltonian describing particle–hole creation can be written in the second quantized form as [9,10]

$$\hat{H}_0 = \sum_k \varepsilon_k^e \hat{c}_k^\dagger \hat{c}_k + \sum_k \varepsilon_k^h \hat{d}_k^\dagger \hat{d}_k \ . \tag{1}$$

Here, $\hat{c}_k^\dagger$ and $\hat{c}_k$ are creation and annihilation operators for electrons, while $\hat{d}_k^\dagger$ and $\hat{d}_k$ are the corresponding operators for holes. $\varepsilon_k^e$ and $\varepsilon_k^h$ represent the respective energy dispersions for electrons and holes in their bands where $\varepsilon_k^h = -\varepsilon_k^e$ at time t=0. The system remains symmetric in the absence of external interactions, as the creation of an electron is always accompanied by the creation of a hole.

In economic terms, this means that loan issuance generates a balanced money–debt pair without inherent gain or loss. Money, analogous to a quasiparticle, flows through the economy; debt, analogous to a hole, remains stationary in the institutional field. To model economic asymmetry caused by time-dependent external forces such as profit and interest [2,5], we introduce perturbation terms as

$$\hat{V}(t) = \sum_k \left[ \Pi_k(t)\hat{c}_k^\dagger \hat{c}_k + r_k(t)\hat{d}_k^\dagger \hat{d}_k \right], \qquad (2)$$

where $\Pi_k(t)$ is a time-dependent profit rate acting on monetary states $(c^k \hat{c}_k)$ and $r_k(t)$ is the time-dependent interest rate acting on debt states $(d^k \hat{d}_k)$. The initial condition must obey $\hat{V}(t=0) = 0$. These perturbations shift the energies of the money and debt components asymmetrically [2,5], resulting in a divergence in their respective evolutions.

To further this analogy, the creditor–debtor relationship can be viewed as a bound state, similar to an exciton in condensed matter systems [9,10]. The quasiparticle (money) and quasihole (debt) are not merely created, but also coupled through contractual obligations and policy-imposed potentials (e.g., repayment schedules, interest terms). This binding introduces structure and risk to the economic exciton. In special cases such as quantitative easing (QE) [1,2], the simultaneous creation of central bank money and government bonds leads to a coupled state that exhibits systemic correlation. This behavior closely parallels quantum entanglement [11,12], wherein measuring one part of a coupled state instantly constrains the behavior of the other. As long as there is no external economic field (e.g., interest rate or investment return), the money–debt pair is symmetric and can annihilate via repayment. However, once time-dependent perturbations such as profit $\Pi_k(t)$ or interest $r_k(t)$ are introduced, the symmetry is broken, and the evolution of money and debt diverges, leading to accumulation or imbalance [2,5]. This sets the stage for modeling financial systems as dynamic fields where time-dependent interactions define systemic outcomes and where exciton-like binding and entangled responses shape economic behavior.

## 3. Mechanisms of Pair Creation: QE and Asset-Backed Creation

This section compares two distinct methods of money creation—Quantitative Easing (QE) [1,2] and gold-backed issuance [3]—through the lens of quantum field theory [9-12]. Both mechanisms result in the creation of money, but they differ fundamentally in how the accompanying asset behaves and how symmetry is preserved or broken. We describe each using quantum states, operators, and spin representations, and summarize the conceptual differences in a comparative table.

### 3.1 QE-type Creation

In fiat-based monetary systems, particularly under quantitative easing (QE) [1,2], money is created through central bank operations that purchase government bonds using newly issued currency. This process generates a pair: money that enters the economy, and a bond liability that resides on the central bank's balance sheet. The two are tightly linked in function and time of creation, though spatially separated in the economic field.

This pair creation may be modeled as

$$|\text{QE Pair}\rangle = |\text{Money}^\uparrow, \text{Bond}^\downarrow\rangle. \quad (3)$$

- $|\text{Money}^\uparrow\rangle$ represents fiat money as a spin-up excitation (electron)

- $|\text{Bond}^\downarrow\rangle$ represents the bond as a spin-down hole-like state

The process resembles electron–hole creation in a vacuum field. The central bank injects money ($\hat{c}_k^\dagger$) and absorbs a bond ($\hat{d}_k^\dagger$):

$$\hat{H}_{\text{QE}} = \sum_k \left( \hat{c}_k^\dagger \hat{d}_k^\dagger + \text{h.c.} \right). \quad (4)$$

These states are entangled in economic behavior—disturbance to one (e.g., bond selloff) influences the stability or valuation of the other (e.g., money confidence).

### 3.2 Asset-Backed Creation

In a gold-backed monetary system [3], money is created not by acquiring debt, but by referencing existing tangible assets such as gold. The gold remains physically stored at the central bank and serves as a claimable anchor for the issued money.

In this model, each unit of money exists as a superposition of gold and fiat components

$$|\text{Asset}\rangle = a|\text{Money}^\uparrow\rangle + b|\text{Gold}^\downarrow\rangle. \quad (5)$$

- This implies that each asset behaves as a single quantum particle (quasiparticle) with two accessible spin states—spin-up representing fiat money and spin-down representing gold. The distinction arises because gold is not the same species of asset as fiat money; hence, the spin-down designation is chosen to signify its complementary state.

- The system allows reversible transformation: use of money suppresses gold activity, and redeeming gold annihilates its fiat counterpart.

We define a spin-flip operation that switches between the two basis states [13]

$$\hat{\sigma}_x |\text{Gold}^\downarrow\rangle = |\text{Money}^\uparrow\rangle, \quad \hat{\sigma}_x |\text{Money}^\uparrow\rangle = |\text{Gold}^\downarrow\rangle , \tag{6}$$

where $\sigma_x$ is Pauli spin matrix. Unlike QE, no entangled pair is created; instead, the total system preserves symmetry through confinement of gold as a static vacuum field.

These two mechanisms highlight different philosophical underpinnings (see Table 1): QE generates liquidity through synthetic pairing of financial instruments, while asset-backed systems preserve underlying value by embedding monetary dynamics in a fixed resource reference. This field-theoretic model reveals the deep symmetry logic embedded within monetary creation frameworks [1-3].

**Table 1**: Comparison Between QE and Gold-Backed Monetary Creation

| Property | QE-Type Creation | Asset-Backed Creation |
| --- | --- | --- |
| Primary-pair created | Money ↑, Bond ↓ | Superposition: Money ↑, Gold ↓ |
| Quantum structure | Entangled pairs | Coherent superposition |
| Convertibility | No direct convertibility | Reversible via $\sigma_x$ operation |
| Asset mobility | Bonds may circulate in markets | Gold remains confined in central vault |
| Symmetry structure | Broken symmetry via interest/market perturbations | Maintained symmetry with potential-like gold backing |
| Field interpretation | Quasiparticle–quasihole in monetary vacuum | Superposition collapse based on usage |

## 4. Discussion

This model offers a field-theoretic view of monetary dynamics that can be applied to analyze:

- Systemic risk accumulation from asymmetrical pair creation (e.g., high-interest credit bubbles)
- Stability in asset-backed systems compared to pure credit expansion
- Potential redefinition of monetary policies in terms of symmetry conservation It also opens interdisciplinary dialogue between economics, physics, and information systems.

A particularly notable implication of this framework is the strong systemic correlation between money and bonds created via quantitative easing (QE). When central banks engage in QE, they generate new money by purchasing bonds—establishing a coupled relationship between currency issuance and bond holdings[1,2]. While not a literal quantum entanglement, this coupling exhibits features analogous to entangled states in physics: simultaneous creation, mutual dependence, and immediate feedback to perturbations.

For instance, when U.S. Treasury bonds are massively sold off in secondary markets, it affects the perceived stability of the U.S. dollar almost instantaneously. This behavior resembles entangled-like responses: just as measuring one particle of an entangled pair in quantum mechanics instantaneously affects the other, financial shocks to bond markets propagate through institutional channels and market psychology, influencing the value and function of the money created alongside those bonds.

This relationship can be conceptually represented as a correlated two-state system. In particular, we propose a Bell-state-like entangled configuration to describe QE-based monetary dynamics [11,12]:

$$|\Phi\rangle = \frac{1}{\sqrt{2}}\left(|\text{Money}^\downarrow, \text{Bond}^\uparrow\rangle + |\text{Money}^\uparrow, \text{Bond}^\downarrow\rangle\right) \qquad (7)$$

Here, $|\text{Money}^\downarrow\rangle$ represents a depreciating currency, while $|\text{Money}^\uparrow\rangle$ denotes an appreciating one. Likewise, $|\text{Bond}^\uparrow\rangle$ indicates a rising yield (falling bond price), and $|\text{Bond}^\downarrow\rangle$ a falling yield (rising bond price). This entangled-state formulation captures the reciprocal nature of market perception: a sharp sell-off in bonds ($\text{Bond}^\uparrow$) corresponds to downward pressure on money ($\text{Money}^\downarrow$), while rising confidence in currency ($\text{Money}^\uparrow$) often supports bond markets ($\text{Bond}^\downarrow$). Disturbance to the bond leg (e.g., rapid selloff or revaluation) causes a reconfiguration in the monetary counterpart. This concept emphasizes that QE not only increases liquidity but also embeds systemic interdependence between financial instruments that may be spatially or institutionally separated.

Although economic literature typically avoids the language of quantum systems, studies from the Bank of England [2] and works such as Werner (2014) [5] and McLeay et al. (2014) [2] support the view that money created through QE is intrinsically linked to the liabilities and financial instruments used to generate it. These insights justify the use of an entanglement-like analogy to capture the depth of systemic co-dependence introduced by QE.

### 4.1 QE Pairs as Bell-Type Entangled States: An Analogy

To deepen the physical interpretation of QE, we propose modeling the money–bond relationship as a Bell-type entangled state. This representation reflects the dual, mutually dependent nature of financial perception in QE-based systems:

$$|\Phi_{QE}\rangle = \frac{1}{\sqrt{2}}\left(|\text{Money (weakened)}, \text{Bond (yield}\uparrow)\rangle + |\text{Money (strengthened)}, \text{Bond (yield}\downarrow)\rangle\right) \cdot \qquad (8)$$

This quantum analogy encodes the idea that market confidence or disruption in one component (e.g., bonds) instantaneously influences the behavior and value of the entangled counterpart (e.g., currency). Such feedback mechanisms are consistent with observed financial reactions:

- When bonds are sold off and yields rise, the associated fiat currency tends to weaken.

- When demand for fiat currency increases, bond prices often rise and yields fall, reinforcing the system's liquidity and perceived stability.

From a field-theoretic perspective, the QE entangled pair is not spatially colocated but dynamically correlated. Their joint state cannot be described as the product of independent components:

$$\rho_{QE} \neq \rho_{Money} \otimes \rho_{Bond} \quad . \tag{9}$$

This reflects the systemic coupling that QE embeds within the macroeconomic field. Using this framework, future monetary analysis can incorporate symmetry-breaking, decoherence effects, and policy-induced collapse of financial wavefunctions. The Bell state becomes a compact yet powerful metaphor to understand the sensitivity of the financial system to perturbations introduced at any entangled node.

**4.2 Microeconomic Loan Creation as Localized Entangled Pair** At the individual level, borrowing from a bank can also be viewed as a form of localized entangled pair creation. When a person takes out a loan:

- The money enters circulation (e.g., to buy goods or services)
- The corresponding debt is stored in the banking system

This results in a micro-level asymmetric pair:

$$|\psi_{loan}\rangle = |Money^{\uparrow}_{released}, Debt^{\downarrow}_{stored}\rangle \quad . \tag{10}$$

Here, only the money is observable and mobile within the economy, while the debt remains as a confined, invisible counterpart recorded as a liability. Repayment of the loan corresponds to a form of recombination that collapses the pair and restores local monetary symmetry.

This structure is reminiscent of the creation of **an exciton** in condensed matter physics, where an electron is excited and leaves behind a positively charged hole. Just as the exciton represents a bound state of electron–hole pairs influencing the electronic properties of materials, a loan represents **a bound money–debt pair** influencing the liquidity and risk profile of the economic system.

**4.3 Comparative Structure of Money with and without Paired Debt**

We distinguish two fundamental types of money based on their origin and entangled counterpart as is given in table 2. Only credit money results in a quasiparticle–quasihole pair. Earned or asset-exchange money emerges without introducing residual economic tension, leading to more stable systemic behavior.

**Table 2** comparison between Money with and without Paired Debt

| Type of Money | Origin | Associated Debt? | Quantum Model | Economic Implication |
|---|---|---|---|---|
| Credit Money | Loan from Bank | Yes | |Money↑, Debt↓⟩ | Creates systemic risk due to outstanding liability |
| Earned/Asset Money | Wages or Asset Sale | No | |Money↑⟩ | Clean contribution to circulation without backpressure |

To further develop the analogy between credit-money pairs and physical excitons, we may describe the interaction using a Schrödinger-like framework. Let $\psi(x)$ represent the spatial probability amplitude of a money–debt pair (economic exciton) in economic configuration space x. The economic potential $V(x)$ represents constraints, interest rates, or collateral conditions imposed by institutions:

$$\left[-\frac{\hbar^2}{2m}\nabla^2 + V(x)\right]\psi(x) = E\psi(x) \quad . \tag{11}$$

Alternatively, in a second quantization framework, the creation of a money–debt pair (analogous to **an exciton**) is described by [9,10]

$$\hat{H}_{\text{econ-exciton}} = \sum_k \epsilon_k^M \hat{c}_k^\dagger \hat{c}_k + \sum_q \epsilon_q^D \hat{d}_q^\dagger \hat{d}_q + \sum_{k,q} U_{kq} \hat{c}_k^\dagger \hat{d}_q^\dagger + \text{h.c.} \quad . \tag{12}$$

Here:

- $\hat{c}_k^\dagger$: creates a unit of money (electron-like excitation)

- $\hat{d}_q^\dagger$: creates a debt record (hole-like excitation)

- $U_{kq}$: encodes institutional or policy-driven coupling between money and debt

- The last term represents their correlated creation, similar to exciton formation

This model lays the groundwork for describing the evolution, recombination, and annihilation of financial excitons within the macroeconomic field.

**4.4 Asset Interchange as Lattice Operator on Economic Sites**

In market-based exchanges (such as buying/selling goods), money and assets are not created or destroyed but simply transferred between agents. This resembles an interchange of particle states across economic lattice sites i and j [10, 13]:

$$\hat{P}_{ij} | \text{Asset}_i, \text{Empty}_j \rangle = | \text{Empty}_i, \text{Asset}_j \rangle \quad . \tag{13}$$

This operator $\hat{P}_{ij}$ describes transactions like:

- $\hat{P}_{ij} | \text{Money}_A, \text{Product}_B \rangle \rightarrow | \text{Product}_A, \text{Money}_B \rangle \quad .$ (14)

Such exchanges conserve overall value and energy in the system and do not introduce holes or credit burdens. They are essential to maintaining dynamic equilibrium in economic fields, in contrast to asymmetrical pair creation mechanisms like loans or QE.

### 4.5 Informal Lending and Violence: Short-Range Interactions in Economic Excitons

In contrast to formal financial systems—where creditor and debtor are separated by legal structures, regulatory buffers, and institutional oversight—**informal lending** (or "loan shark" credit) exists in an unregulated domain where physical, social, and psychological interactions dominate. This environment gives rise to **short-range, high-intensity interactions** between money and debt, often manifesting as coercion, threat, or violence. Such dynamics can be modeled using a field-theoretic Hamiltonian augmented with violation-type interaction terms.

The emergence of money–debt pairs in this context occurs not in well-regulated institutional environments but within **economic vacuums**—regions of the system characterized by extreme disparities between the rich and the poor. Individuals who participate in informal lending typically **lack access to formal credit**, meaning they reside in economic regions with **no credit potential or collateral**, i.e., deep financial gaps. In contrast to banking systems where credit is mediated by financial infrastructure, in informal settings, **pair creation arises spontaneously across the economic band gap** separating wealth classes.

We propose the following second-quantized Hamiltonian for informal lending systems:

$$\hat{H}_{\text{informal}} = \hat{H}_{\text{free}} + \hat{H}_{\text{binding}} + \hat{H}_{\text{viol}} + \hat{H}_{\text{asym}}(t) \quad , \tag{15}$$

Where each term is defined as follows:

**1. Free Particle Terms:**

$$\hat{H}_{\text{free}} = \sum_k \epsilon_k^M \hat{c}_k^\dagger \hat{c}_k + \sum_q \epsilon_q^D \hat{d}_q^\dagger \hat{d}_q \quad . \tag{16}$$

- $\hat{c}_k^\dagger, \hat{c}_k$ : creation/annihilation operators for money.

- $\hat{d}_q^\dagger, \hat{d}_q$ : creation/annihilation operators for debt.

- $\epsilon^M, \epsilon^D$ : represent liquidity potential and debt burden.

**2. Binding Term (Exciton Creation):**

$$\hat{H}_{binding} = \sum_{k,q} U_{kq} \hat{c}_k^\dagger \hat{d}_q^\dagger + \text{h.c.} \quad . \tag{17}$$

- Represents the issuance of a loan or credit contract, pairing money and debt.

**3. Short-Range Coulomb-like Interaction:**

$$\hat{H}_{viol} = \sum_{k,q} V_{kq}^{viol} \hat{c}_k^\dagger \hat{c}_k \hat{d}_q^\dagger \hat{d}_q \quad . \tag{18}$$

- This term replaces the traditional Coulomb interaction. It reflects **unscreened, coercive social interactions** that occur in informal lending: violence, threats, or immediate enforcement.
- While structurally similar to a Coulomb interaction, this term captures **economic violations** arising in the absence of legal/institutional screening.
- We assume $V_{kq}^{Viol} \propto f(\Delta_{pr})$, where $\Delta_{pr}$ is the gap between poor and rich, i.e., the severity of economic disparity.

**4. Time-dependent Asymmetry (Profit and Interest):**

$$\hat{H}_{asym}(t) = \sum_k \delta\epsilon^M(t) \hat{c}_k^\dagger \hat{c}_k + \sum_q \delta\epsilon^D(t) \hat{d}_q^\dagger \hat{d}_q . \tag{19}$$

- $\delta\epsilon^M(t) = \Pi_k(t)$: economic gain or profit from lending (affecting money evolution).

- $\delta\epsilon^D(t) = r_k(t)$: compounding interest or escalating penalty (affecting debt).

---

**Interpretation:** This full Hamiltonian captures both the **creation** and **enforcement** mechanisms of debt in unregulated contexts. The violation-type term $\hat{H}_{viol}$ models physical or social coercion, especially prevalent where legal institutions are absent or ineffective. **The existence of money–debt excitons in the informal sector signifies the presence of individuals excluded from the formal credit system—thus, the total number of such excitons is a direct proxy for the** **population of the poor.** Moreover, the **strength and range of violation-type interactions**—reflected in $V^{viol}$—is directly correlated with the economic gap between rich and poor.

**In systems with extreme inequality, these interaction terms become dominant**, leading to greater instability and higher likelihood of violence. This approach suggests that the measurement of informal credit pair density and interaction strength could serve as a novel metric for diagnosing poverty concentration and systemic social risk. **Theoretical Context and Literature Basis:** The structure of the Hamiltonian used in this section draws formal inspiration from excitonic models in condensed matter physics, particularly those incorporating interaction terms such as short-range Coulomb interactions and mean-field approximations. The binding term $\hat{H}_{binding}$ is analogous to exciton formation due to attractive electron–hole coupling, while the violation-like term $\hat{H}_{viol}$ serves as an economic analog to unscreened, local interactions—structurally reminiscent of effective interaction terms used in models of strongly correlated systems [9,10]. However, instead of electromagnetic interactions, these terms encode coercive enforcement, social pressure, and risk in informal economic contexts. This approach also parallels frameworks in econophysics that attempt to use field-theoretic representations of economic risk, such as in the modeling of systemic fragility and contagion in complex financial networks. Here, we propose that short-range, high-intensity interactions between money and debt—especially in economically distressed regions—can be described using a violation-based potential dependent on socioeconomic disparity, similar in structure but distinct in meaning from traditional Coulombic potentials.

## 5. Summary

This work presents a novel interdisciplinary framework that models monetary creation and credit mechanisms using principles from quantum field theory and condensed matter physics [9-12]. By drawing a direct analogy between particle–hole excitations in a Fermi Sea and the creation of money–debt pairs in economic systems [2,5], we establish a consistent formalism for analyzing the structure, evolution, and stability of financial processes.

We demonstrate that:

- **Quantitative Easing (QE)** creates entangled money–bond pairs that resemble quasiparticle–quasihole excitations in a vacuum [1,2,9,10]. This process introduces systemic coupling between financial instruments and exhibits symmetry-breaking under the influence of external economic forces, such as interest rates and market expectations [4,5].

- **Asset-backed money** operates analogously to quantum superposition, where the monetary state may collapse into either fiat or gold depending on usage, with a reversible structure and preserved symmetry [3,10].

14- **Microeconomic credit creation**, such as individual bank loans, corresponds to localized exciton-like states [9,10]. These money–debt pairs can be modeled using second quantization formalism and perturbed Hamiltonians [11,12] to capture asymmetric time evolution influenced by profit and interest [2,5].

- **Earned income or asset-based transactions**, in contrast, do not generate hole-like debt states. These transactions can be represented as single-particle wavefunctions that circulate without generating systemic backpressure [4,8], supporting long-term equilibrium.

- **Market-based asset exchanges** are described as lattice-site interchanges without net creation or destruction [10], further reinforcing the utility of field-based analogies for economic modeling.

- Informal lending systems—emerging outside institutional frameworks—are modeled as excitonic money–debt pairs forming in economic vacuum regions, where the credit gap between the rich and poor is widest. Unlike regulated credit creation, these informal excitons involve short-range, coercive interactions that replace legal enforcement. The model introduces violation-type interaction terms in the Hamiltonian in section 4.5 to represent this unregulated force, structurally resembling unscreened Coulomb interactions in condensed matter systems. The number of such pairs can serve as a proxy for poverty levels, while the strength of interaction reflects the magnitude of socioeconomic inequality—linking economic disparity to systemic fragility and social violence.

Finally, the analogy of **financial excitons**—bound pairs of credit and debt—provides insight into risk propagation and systemic fragility [5,9,10]. This approach opens a new direction for econophysics: one grounded not only in statistical models or thermodynamics, but in **field-theoretic and quantum-coherent interpretations** of economic structure [10-12].

**References**

[1] Bernanke, B. S. (2020). The New Tools of Monetary Policy. *American Economic Review*, 110(4), 943–983.

[2] McLeay, M., Radia, A., & Thomas, R. (2014). Money Creation in the Modern Economy. *Bank of England Quarterly Bulletin*, Q1 2014, 14–27.

[3] Bordo, M. D. (1984). *The Gold Standard: The Traditional Approach*. In M. D. Bordo & A. J. Schwartz (Eds.), *A Retrospective on the Classical Gold Standard, 1821–1931* (pp. 23–113). University of Chicago Press.

[4] Mishkin, F. S. (2019). *The Economics of Money, Banking, and Financial Markets* (11th ed.). Pearson.

[5] Werner, R. A. (2014). Can Banks Individually Create Money Out of Nothing? – The Theories and the Empirical Evidence. *International Review of Financial Analysis*, 36, 1–19.